\documentclass[AMA,Times1COL]{WileyNJDv5} 

\usepackage[binary-units=true]{siunitx}
\DeclareSIUnit{\bph}{\textup{bph}}
\usepackage{amsmath}
\usepackage{amsfonts}%
\usepackage{amssymb}%
\usepackage[final]{changes}
\usepackage{graphicx}
\usepackage{hyperref}
\usepackage{amssymb}

\usepackage{tikz,pgfplots}
\usepackage{transparent} 
\usepackage{subfigure} 

\usepgfplotslibrary{fillbetween}
\usetikzlibrary{plotmarks}
\usetikzlibrary{decorations.pathreplacing}
\usetikzlibrary{babel}
\usetikzlibrary{calc}
\usetikzlibrary{arrows}
\usetikzlibrary{patterns}
\usetikzlibrary{shadows}
\pgfplotsset{compat=1.10}
\usepackage{rotating}
\usepackage{array}
\usepackage{tabulary}
\usepackage{multirow}
\usepackage{booktabs}
\usepackage{colortbl}
\usepackage{xparse,xspace}

\usepackage{amsmath}%
\usepackage{amsfonts}%
\usepackage{amssymb}%
\usepackage{xcolor}

\usepackage{textcomp}

\definecolor{cinnabar}{rgb}{0.89, 0.26, 0.2}

\definechangesauthor[name={Laurent}, color=blue]{ld}
\definechangesauthor[name={Fred}, color=magenta]{fp}
\definechangesauthor[name={Fred}, color=red]{fp2}
\definechangesauthor[name={Lauriane}, color=orange]{lb}
\definechangesauthor[name={Christophe}, color=blue]{cp}

\newcommand{\fpd}[1]{\deleted[id=fp]{}}
\newcommand{\fpx}[1]{\added[id=fp2]{[check] #1}}

\definecolor{IFPENBlue}{rgb}{0, 0.42, 0.7}

\definecolor{IFPENBluelgt}{rgb}{0, 0.52, 0.8}

\definecolor{IFPENBluelgtlgt}{rgb}{0, 0.62, 0.9}

\definecolor{IFPENGreen}{rgb}{0.5412,    0.7294,    0.0902}

\definecolor{cinnabar}{rgb}{0.89, 0.26, 0.2}
\definecolor{darkcoral}{rgb}{0.8, 0.36, 0.27}
\definecolor{darkelectricblue}{rgb}{0.33, 0.41, 0.47}
\definecolor{darkcyan}{rgb}{0.0, 0.55, 0.55}
\definecolor{dark-red}{rgb}{0.6,0,0}
\definecolor{darkchestnut}{rgb}{0.6, 0.41, 0.38}

\definecolor{jasper}{rgb}{0.84, 0.23, 0.24}
\definecolor{indianred}{rgb}{0.8, 0.36, 0.36}
\definecolor{fireenginered}{rgb}{0.81, 0.09, 0.13}
\definecolor{darkpastelred}{rgb}{0.76, 0.23, 0.13}
\definecolor{brickred}{rgb}{0.8, 0.25, 0.33}
\definecolor{greenmunsell}{rgb}{0.0, 0.66, 0.47}
\definecolor{lightseagreen}{rgb}{0.13, 0.7, 0.67}
\definecolor{tealgreen}{rgb}{0.0, 0.51, 0.5}
\definecolor{lightslategray}{rgb}{0.47, 0.53, 0.6}
\definecolor{lightseagreen}{rgb}{0.13, 0.7, 0.67}
\definecolor{deepcarrotorange}{rgb}{0.91, 0.41, 0.17}
\definecolor{deepcarminepink}{rgb}{0.94, 0.19, 0.22}
\definecolor{deepcarmine}{rgb}{0.66, 0.13, 0.24}
\definecolor{rufous}{rgb}{0.66, 0.11, 0.03}
\definecolor{sacramentostategreen}{rgb}{0.0, 0.34, 0.25}
\definecolor{carminepink}{rgb}{0.92, 0.3, 0.26}

\definecolor{lightslategray}{rgb}{0.47, 0.53, 0.6}
\definecolor{blue1}{rgb}{0.01, 0.31, 0.59}
\definecolor{blue2}{rgb}{0.01, 0.31, 0.59}
\definecolor{blue3}{rgb}{0.0, 0.47, 0.75}
\definecolor{blue4}{rgb}{0.29, 0.59, 0.82}

\definecolor{darkgray}{rgb}{0.6, 0.6, 0.6}

\definecolor{gray}{rgb}{0.8, 0.8, 0.8}

\definecolor{darkdarkgray}{rgb}{0.4, 0.4, 0.4}

\NewDocumentCommand {\msh}{o o}{%
	\IfNoValueTF{#2}{
		\IfNoValueTF{#1}{\ensuremath{Ms}\xspace}
		{\ensuremath{Ms(#1)}\xspace}
	}
	{\ensuremath{Ms_{#2}(#1)}\xspace}
	\xspace}

\NewDocumentCommand \mshh {o o}{%
	\IfNoValueTF{#2}{
		\IfNoValueTF{#1}{\ensuremath{\widehat{Ms}}\xspace}
		{\ensuremath{\widehat{Ms}(#1)}}
	}
	{\ensuremath{\widehat{Ms}_{#2}(#1)}}
}
\NewDocumentCommand \mshcp {o o}{%
	\IfNoValueTF{#2}{
		\IfNoValueTF{#1}{\ensuremath{\overset{*}{Ms}}}
		{\ensuremath{\overset{*}{Ms}(#1)}}
	}
	{\ensuremath{\overset{*}{Ms}_{#2}(#1)}}
}

\newcommand{\hs}{Hexa\-Shrink\xspace}
\newcommand{\lund}{{{\rmfamily\protect\scshape {Lundi}}\textsubscript{sim}}\xspace}

\newcommand{\nsZ}{\ensuremath{\mathrm{nearshore_{0}}}\xspace}
\newcommand{\nsO}{\ensuremath{\mathrm{nearshore_{1}}}\xspace}
\newcommand{\nsA}{\ensuremath{\mathrm{nearshore_{a}}}\xspace}
\newcommand{\flv}{\ensuremath{\mathrm{fluvial}}\xspace}

\NewDocumentCommand {\txtpr}{o o}{%
	\IfNoValueTF{#2}{
		\IfNoValueTF{#1}{$P$\hspace{-6pt}}
		{$P(#1)$\hspace{-6pt}}
	}
	{$P_{#2}(#1)$\hspace{-6pt}}
	\xspace}

\NewDocumentCommand \txtprh {o o}{%
	\IfNoValueTF{#2}{
		\IfNoValueTF{#1}{$\widehat{P}$\hspace{-6pt}}
		{$\widehat{P}(#1)$\hspace{-6pt}}
	}
	{$\widehat{P}_{#2}(#1)$\hspace{-6pt}}
}

\NewDocumentCommand {\pr}{o o}{%
	\IfNoValueTF{#2}{
		\IfNoValueTF{#1}{\ensuremath{P}\xspace}
		{\ensuremath{P(#1)}\xspace}
	}
	{\ensuremath{P_{#2}(#1)}\xspace}
	\xspace}

\NewDocumentCommand \prcp {o o}{%
	\IfNoValueTF{#2}{
		\IfNoValueTF{#1}{\ensuremath{\overset{*}{P}}}
		{\ensuremath{\overset{*}{P}(#1)}}
	}
	{\ensuremath{\overset{*}{P}_{#2}(#1)}}
}

\NewDocumentCommand \prh {o o}{%
	\IfNoValueTF{#2}{
		\IfNoValueTF{#1}{\ensuremath{\hat{P}}\xspace}
		{\ensuremath{\hat{P}(#1)}}
	}
	{\ensuremath{\hat{P}_{#2}(#1)}}
}

\NewDocumentCommand \cell {o}{%
	\IfNoValueTF{#1}{\ensuremath{\mathcal{C}}\xspace}
	{\ensuremath{\mathcal{C}(#1)}}
}

\NewDocumentCommand \Cmp {o o}{%
	\IfNoValueTF{#2}{
		\IfNoValueTF{#1}{\ensuremath{\Lambda}}
		{\ensuremath{\Lambda_{#1}}}
	}
	{\ensuremath{\Lambda_{#1}(#2)}}
}	

\NewDocumentCommand \mathCmp {o o}{%
	\IfNoValueTF{#2}{
		\IfNoValueTF{#1}{\ensuremath{\Lambda}}
		{\ensuremath{\Lambda_{#1}}}
	}
	{\ensuremath{\Lambda_{#1}(#2)}}
}

\NewDocumentCommand \CmpInv {o o}{%
	\IfNoValueTF{#2}{
		\IfNoValueTF{#1}{\rotatebox[origin=c]{180}{\ensuremath{\Lambda}\xspace}}
		{\rotatebox[origin=c]{180}{\ensuremath{\Lambda}}\ensuremath{_{#1}}}
	}
	{\rotatebox[origin=c]{180}{\ensuremath{\Lambda}}\ensuremath{_{#1}(#2)}}
}

\NewDocumentCommand \mathCmpInv {o o}{%
	\IfNoValueTF{#2}{
		\IfNoValueTF{#1}{\rotatebox[origin=c]{180}{\ensuremath{\Lambda}\xspace}}
		{\rotatebox[origin=c]{180}{\ensuremath{\Lambda}}\ensuremath{_{#1}}}
	}
	{\rotatebox[origin=c]{180}{\ensuremath{\Lambda}}\ensuremath{_{#1}(#2)}}
}

\NewDocumentCommand \cmp {o o}{%
	\IfNoValueTF{#2}{
		\IfNoValueTF{#1}{\ensuremath{\lambda}}
		{\ensuremath{\lambda_{#1}}}
	}
	{\ensuremath{\lambda_{#1}(#2)}}
}

\NewDocumentCommand \cmpInv {o o}{%
	\IfNoValueTF{#2}{
		\IfNoValueTF{#1}{\rotatebox[origin=c]{180}{\ensuremath{\lambda}\xspace}}
		{\rotatebox[origin=c]{180}{\ensuremath{\lambda}}_{#1}}
	}
	{\rotatebox[origin=c]{180}{\ensuremath{\lambda}}\ensuremath{_{#1}(#2)}}
}

\DeclareSIUnit\darcy{d}
\DeclareSIUnit\poise{P} 
\articletype{Data Paper}%

\received{Date Month Year}
\revised{Date Month Year}
\accepted{Date Month Year}
\journal{Journal}
\volume{00}
\copyyear{2024}
\startpage{1}

\begin{document}

\title{\lund:  model meshes for flow simulation and scientific data compression benchmarks (PREPRINT)}

\author[1]{Laurent Duval}
\author[2]{Frédéric Payan}
\author[1]{Christophe Preux}
\author[1,2]{Lauriane Bouard}

\authormark{Duval \textsc{et al.}}
\titlemark{\lund:  model meshes for flow simulation and scientific data compression benchmarks}
\address[1]{\orgdiv{IFP Energies nouvelles}, \orgaddress{\country{France}}}
\address[2]{\orgdiv{Université Côte d’Azur}, \orgname{CNRS, I3S}, \orgaddress{\country{France}}}

\corres{Corresponding author Laurent Duval \email{laurent.duval@ifpen.fr}}



\abstract[Abstract]{The volume of scientific data  produced for and by numerical simulation workflows is increasing at an incredible rate. This raises concerns either in computability, interpretability, and sustainability. This is especially noticeable  in  earth science (geology, meteorology, oceanography, and astronomy), notably with climate studies. 	
	We highlight five main evaluation  issues: efficiency, discrepancy, diversity, interpretability, availability. 
	Among  remedies, lossless and lossy compression techniques are becoming popular to better manage dataset volumes. Performance assessment --- with comparative benchmarks --- require open datasets shared under FAIR principles  (Findable, Accessible, Interoperable, Reusable), with MRE (Minimal Reproducible  Example) ancillary data for reuse.
We share   \lund, an exemplary  faulted geological mesh. It is inspired by SPE10 comparative Challenge.
Enhanced by porosity/permeability  datasets, this dataset proposes four distinct subsurface environments. They were primarily designed for flow simulation in porous media. Several consistent  resolutions (with  HexaShrink multiscale representations) are proposed for each model. We also provide a set of reservoir features  for reproducing typical two-phase flow simulations on all \lund models in a  reservoir engineering context. This dataset is chiefly meant for benchmarking and evaluating data size reduction (upscaling) or genuine composite mesh compression algorithms. It is also suitable for other advanced mesh processing workflows in geology and reservoir engineering, from  visualization to machine learning.
 \lund meshes are available at  \href{10.5281/zenodo.14641959}{10.5281/zenodo.14641959}.
}

\keywords{compression, flow simulation, geology, open data, reservoir engineering,  scientific data,  simulation, volumic model mesh\\
	An enhanced revision of this preprint has been accepted to: Geoscience Data Journal, Royal Meteorological Society (RMetS), Wiley (August 2025). The published paper will be available at \href{http://doi.org/10.1002/gdj3.70030}{10.1002/gdj3.70030}}

\jnlcitation{\cname{%
\author{Duval L},
\author{Payan F},
\author{Preux C}, and
\author{Bouard L}}.
\ctitle{\lund:  model meshes for flow simulation and compression benchmarks.} \cjournal{\it \fpx{J Comput Phys.}} \cvol{2021;00(00):1--18}.}

\maketitle

\renewcommand\thefootnote{}
\footnotetext{\textbf{Abbreviations:} FAIR, Findable, Accessible, Interoperable, Reusable ; MRE, Minimal Reproducible Example ; HPC, High Performance Computing ; HS, HexaShrink ; CPG, Corner Point Grid.}

\renewcommand\thefootnote{\fnsymbol{footnote}}
\setcounter{footnote}{1}

	\section{Introduction}\label{sec_introduction}
			Science has entered the  ``fourth paradigm'' of data-intensive computing for discovery \cite{Hey_T_2009_book_fourth_pdisd}. 
Increasingly accurate models yield  unprecedented access to more precise simulations, resorting to high-performance computing (HPC) facilities. The exploitation of massive datasets is however  hampered by many size-related  issues, such as storage, memory, workflow management, and visualization \cite{Ahrens_J_2022_j-ieee-comput-graph-appl_technology_tclssv,Sarton_J_2023_j-comput-graph-forum_state-of-the-art_lsvvbsd}, for instance for machine learning and artificial intelligence tasks  \cite{Underwood_R_2024_PREPRINT_understanding_elcmlts}. 	
	As a result, data compression is making a comeback from an influential 1990's multimedia era\footnote{Famous  data compression standards:  jpeg, gif, png, mp3.} to the  many worlds of modeling and simulation. At stake are legal long-term storage issues for instance in climate modeling \cite{Overpeck_J_2011_j-science_climate_dc21c}, checkpoint restart or snapshotting for  fault-tolerance in HPC \cite{Yakushin_I_2020_p-icpp_feature-preserving_lcisda}, approximate computing \cite{Mittal_S_2016_j-acm-comput-surv_survey_tac}, faster selection of  parameters with smaller simulation models, progressive result retrieval \cite{Magri_V_2024_j-ieee-trans-visual-comput-graph_general_fpdcr}, \emph{in situ}/in storage processing \cite{Childs_H_2020_j-int-j-high-perform-comput-appl_terminology_isvas}, objective and subjective performance evaluation, etc. 
	
	For a comprehensive evaluation of compression performance and challenges in modeling \cite{Schweiger_G_2020_j-appl-math-comput_modeling_slssscmp,Underwood_R_2024_PREPRINT_understanding_elcmlts}, several issues deserve attention, sometimes in contrast to what was held true  for multimedia data coding. We thereafter detail  five most prominent issues: efficiency, discrepancy, diversity, interpretability, availability, before we inflect them to geological models in Section \ref{sec_geological_model}, thereby motivating the  \lund mesh dataset \cite{Duval_L_2025_zenodo_lundisim_odmmmpppispe10cgdsa} detailed  in the remaining of the paper. 
	It is available at  \href{10.5281/zenodo.14641959}{10.5281/zenodo.14641959}.
	
	\begin{description}
		\item[Issue 1 (efficiency)]  perfect or lossless  compression like ``zip'' notoriously yields very limited reduction ratios. More than two- to three-fold  reduction in size is rare (except for highly-structured data). Therefore, approximate, near-lossless, progressive or lossy  compression algorithms are required to ensure a significant byte-size reduction compatible with the pressing needs occurring from gigantic simulation volumes. They however entail careful assessments of the data loss impact on performance \cite{Murillo_R_2022_p-annsim_effects_nsa,Taurone_F_2023_incoll_lossless_pfpdec,Walters_M_2023_j-geosci-model-dev_impact_aedpceascmaqm,Wang_D_2024_j-ieee-trans-parallel-distrib-syst_tacp_oeblc3damrs}, especially for bounding errors \cite{Liang_X_2022_j-ieee-tc_magardp_ommebsdr,Liu_J_2024_PREPRINT_cuz-i_hfeblcsdgpus}.
		\item[Issue 2 (discrepancy)]  simulation dataset are typically very heterogeneous. Coming from different sources, at various workflow steps, they include structured, semi-structured, and unstructured data, are of various dimensionality (1D, 2D, 3D and 4D) and stored in various formats (booleans, discrete labels, integers, floats, etc.) or containers (json, HDF, netCDF, FITS, Mexus, ASDF, XLM). They cannot  be addressed with generic compression tools easily, nor with optimal performance. They require  dedicated algorithms, taking into account complicated data morphology \cite{Klower_M_2021_j-nat-comput-sci_compressing_adric}. 
		\item[Issue 3 (diversity)]  mixed and high-dynamics data: each type of data may exhibit a huge diversity of types, ranges, or statistical distributions, from a handful of finite nominal categories (Likert scale, data labels, attributes) to high-precision  values (covering several range scales) with unbalanced histograms \cite{Underwood_R_2023_j-int-j-high-perform-comput-appl_black-box_splcrsd}. At stake here are quantities whose variations have highly non-linear behavior or non-proportional effects. For instance, small values that would be discarded with traditional lossy compression may need to be faithfully preserved.
		\item[Issue 4 (interpretability)] direct interpretation of the different --- and often visually combined --- types of scientific data \cite{Baker_A_2016_j-geosci-model-dev_evaluating_ldccsdle} is less straightforward than with standard audio, image or video \cite{Poppick_A_2020_j-comput-geosci_statistical_alccmd}. First, it is heavily coupled with physical modeling. Second, models potentially undergo  long-lasting simulations whose outputs are subject to a host of objective and subjective assessments. Simulation evaluations gather teams with diverse skills. Their expertise is deployed iteratively, at different stages of the workflow. Owing to simulation  complexity  and compression recency, overall quality assessment is restricted to a small number of individuals from distinct backgrounds, with little universally-accepted metrics and huge policy options. Acceptable objective losses with no influence on simulation may become unacceptable to an expert subjective interpretation, focusing on  specific modalities. In contrast, knowledge of the human sensory systems and the world-wide dissemination of multimedia devices allowed the persistence of widely-accepted compression of audio and visual contents.
		\item[Issue 5 (availability)] open availability of representative models, in FAIR principles \cite{Peters_K_2020_PREPRINT_fair_ltpcessdfrwdccwdcc}, is not granted, for proprietary uses or ad-hoc data manipulation that cannot be reproduced.  One of the co-authors of this paper has for instance encountered apparently huge private meshes which, candidates for potential challenges, showed up to have been artificially inflated by linear interpolation on mid-scale data. This may jeopardize fair compression evaluation, as data becomes highly predictable. Therefore, openly shared geological models, that may be modified to adapt to different simulation contexts, are convenient.
	\end{description} 

\section{Geological model issues}\label{sec_geological_model}
	  
Aside gigantic climate-related models, geoscience somehow lacks open, manageable, heterogeneous data models that can be embedded in a processing or simulation workflow. In a similar way to recent initiatives \cite{Alumbaugh_D_2023_j-geosci-data-j_kimberlina_smdco2mi,Haehnel_P_2023_j-geosci-data-j_development_tdhmingug}, we share with this paper a handful of 3D geological models and their multiscale representations. Developed at IFPEN during Lauriane Bouard's PhD thesis \cite{Bouard_L_2021_phd_refinable_rpvmcsg}, this dataset is collectively denoted as \lund. 
\lund is  dedicated to performance evaluation and benchmarks around the compression of 3D geological models  \cite{Wellmann_F_2018_incoll_3-d_sgmcmu} targeted to simulation workflows, illustrating the five previously raised issues. We named our dataset \lund, after the Icelandic name of the (peaceful) Atlantic puffin. This name is a friendly nod to two protagonistic simulation software suites, Petrel\texttrademark{} and SKUA\texttrademark{}, named after two (highly competing) seabirds. 
\lund was initially created for testing \hs (HS)  \cite{Peyrot_J_2019_j-computat-geosci_hexashrink_pdgpcshmg}, a scalable storage and multiresolution (also called hierarchical \cite{Suter_E_2019_p-eage_principles_hemratffpgr,Abraham_F_2019_j-vis-comput_multiresolution_vmborm,Devarajan_H_2020_p-ieee-ipdps_hcompress_hdcmtse,Ceballos_L_2021_incoll_interactive_dp3dvse}) visualization framework for hexahedral meshes with mixed attributes and discontinuities. HS was then integrated into a comprehensive compression workflow, enabling progressive and refinable data representation of composite hexahedral meshes. ``Composite'' here means that the 3D geometric structure (or grid) may itself be encoded by complementary  spatial locations. In computational geology, this geometry is traditionally structured by a Corner Point Grid (CPG): a 1D coordinate system along the vertical direction (``Pillar'') supports a more horizontal 2D layering (``Zcorn''). This grid may be complemented with numerical properties (porosity, expressed in unit proportion;  permeability given in (milli)darcy or (m)d in the following) and discrete categories (cell activity, rock type) designed from rock physics, for flow simulation in reservoir modeling engineering. 	
	There, a geological model may be filled by different stochastic distributions. They account for  phenomena representing variations in the underground. Once filled with properties, a reservoir model is simulated under varying operating conditions. Such simulations are used to gain insight on how to manage a storage or production facility on a day-to-day basis. 	
	
\begin{figure}[htb]
 	\centering
 	\includegraphics[width=0.6\linewidth]{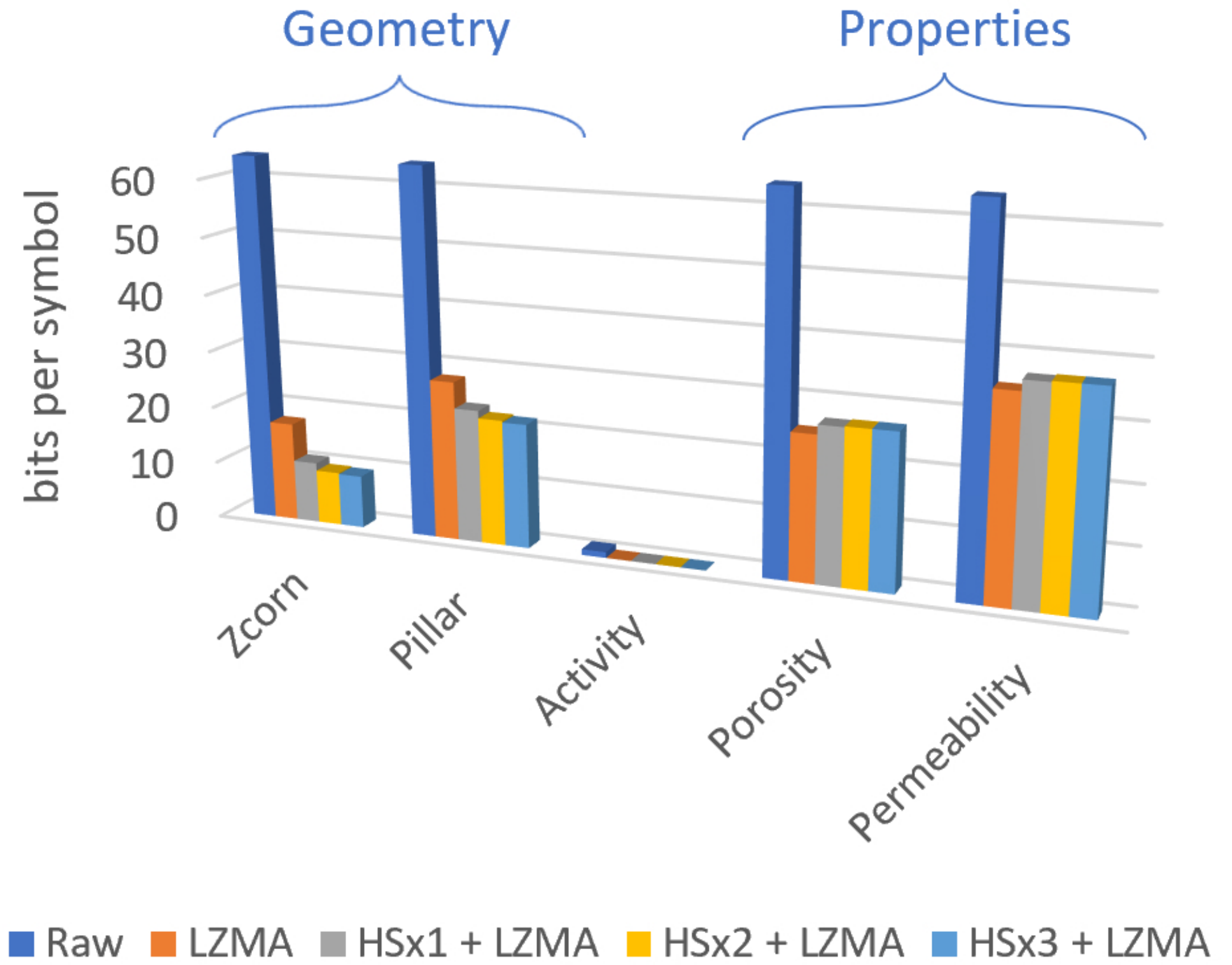}
 	\caption{(Average) number of bits per symbol for each data type of the nearshore$_0$ \lund model (Zcorn, Pillar, Activity, Porosity, Permeability) in  function of the encoding setting: uncompressed raw data (dark blue bars), raw data compressed with LZMA (orange bars), data decomposed at one (gray bars), two (yellow bars) or three (light blue bars) different resolutions before being compressed with LZMA. See \cite[Fig. 17--18]{Peyrot_J_2019_j-computat-geosci_hexashrink_pdgpcshmg} for more visual details.}
 	\label{fig_mesh_object_multiscale}
 \end{figure}

	In  \cite{Peyrot_J_2019_j-computat-geosci_hexashrink_pdgpcshmg}, we observed that different data in  composite meshes distinctly affect compression algorithms. For the sake of completeness, we provide here an illustrative example, based on one of our \lund models, described thereafter. The dark blue bars in Figure \ref{fig_mesh_object_multiscale} represent the ``raw'' number of bits per symbol for various data types in a model cell (i.e., Zcorn, Pillar, Activity, Porosity and Permeability). The orange bar depicts the direct application of the generic yet highly performant lossless LZMA coder (Lempel-Zip-Markov chain Algorithm, \cite[Section 6.26]{Salomon_D_2009_book_handbook_dc})   on all components, with mild average compression (see Issue 1 from Section \ref{sec_introduction}). More specifically, we observe that heterogeneous data types have distinct compression ratios (Issue 2). Boolean Activity property is easily compressed, while Permeability is more challenging. As in \cite{Peyrot_J_2019_j-computat-geosci_hexashrink_pdgpcshmg} we sought at the same time both lossless compression and the possibility to address mesh multiresolution, the gray, yellow and light blue bars of  Figure \ref{fig_mesh_object_multiscale} indicate the LZMA performance after respectively one, two or three levels of multiscale (HS$\times$1, HS$\times$2, HS$\times$3 respectively) decompositions of all properties.

 As seen per the average number of bits per  symbol, integer geometry (Zcorn and Pillar) is increasingly compressed (though mildly) with resolutions, while the effect on continuous scalar properties (Porosity, Permeability) is sightly degraded due to the high-dynamics of the data (Issue 3).
 
Above observations were made on losslessly compressed data. In other terms, decompressed data is faithful to the raw model, hence does not hamper workflow precision, notably in a context of simulation. However, simulation practice often resorts to data at coarser resolutions, for  speedups and multi-scenario evaluations. Plus, it is well-known that different data resolutions (scales) or precision (byte-per-symbol) may subjectively  impact   a simulation workflow (Issue 4). In a typical compress-once/decompress-many context, one may need for instance to address objective mesh size and decompression speed metrics at the beginning of the workflow, and more subjective replays of flow propagation for post-processing.
 Therefore, our \lund dataset contains models at four different levels of resolution to address Issue 5.

The remaining  of the paper is organized as follows. We provide contextual information on reservoir modeling for simulation in  Section \ref{sec_reservoir_modeling_simulation},  inspired from the well-known reservoir engineering challenge SPE10. We craft the two main components of \lund in Section \ref{sec_model_description}: the common model mesh (Subsection \ref{sec_model_description_mesh}) and its  SPE10-inherited physical properties  (Subsection \ref{sec_model_description_properties}). Ancillary data for simulation  are provided in Section \ref{sec_simulation_settings}: global reservoir characteristics (Subsection \ref{sec_global_reservoir_characteristics})  to allow simulation workflow reproduction (MRE, up to software suite characteristics);  the application to fluid production  (Subsection \ref{sec_fluid_production}) with  traditional simulation observables. Section \ref{sec_data_format}  details data availability and associated software. \lund potential reuse and limits  are given in Section \ref{sec_data_reuse}, before  conclusions (Section \ref{sec_data_conclusion}). 

\section{Reservoir modeling for simulation}\label{sec_reservoir_modeling_simulation}

We base our work on a previously published challenge known as SPE10, i.e. the Tenth SPE Comparative Solution Project \cite{Christie_M_2001_j-spe-reserv-eval-eng_tenth_specspcut} for reservoir simulation. We consider its second problem called Model 2, part of the \emph{Brent} sequence (quoting), ``\emph{a waterflood of a large geostatistical model chosen so that it was hard (though not impossible) to compute the true fine-grid solution}'' \cite[p. 308]{Christie_M_2001_j-spe-reserv-eval-eng_tenth_specspcut}. In this challenge, eight companies competed to obtain the best possible outcome in the evaluation of this model, using a combination of simulation software and upscaling techniques. Counter-intuitively (since  data science  is more acquainted with downsampling or downscaling), in the context of reservoir simulation,  upscaling and upgridding denote the process with which a fine-scale geological model (a grid assorted with rock properties such as porosity and permeability data) is converted into coarser models  that are more computationally tractable, while providing outcomes as close as possible as those expected from the finer grid. Cells of the coarser grid (upgridding \cite{King_M_2007_j-ogst_recent_au}) are filled with equivalent properties (upscaling) obtained from  finer-resolution cells, using a variety of homogenization or averaging techniques.  We refer to \cite{Christie_M_2001_j-spe-reserv-eval-eng_tenth_specspcut,Preux_C_2014_j-ogst_use_qirilwpu,Misaghian_N_2018_j-computat-geosci_upscaling_auamruarpa} for  details. 

Upscaling thus reduces the original grid size as well as cell-borne quantities. This results in a global reduction of the size of data with heterogeneous properties, a process similar to what is targeted in genuine data compression, where the modification of data resolution is combined with variations in data precision and additional entropy coding schemes that yield a final compressed file. We refer to \cite{Salomon_D_2009_book_handbook_dc} for advanced notions in data coding.  Meanwhile, one may ask whether suitable data compression, adapted to geological data, is compatible and even maybe beneficial to flow simulation of large heterogeneous models, as partly exposed in  \cite{Bouard_L_2021_phd_refinable_rpvmcsg,Bouard_L_2021_p-coresa_etude_cicpvdsg} (whose outcomes are not required here for further understanding). We now focus  on  \lund benchmark models. 

\section{\lund model description}\label{sec_model_description}

\subsection{{\lund} model mesh}\label{sec_model_description_mesh}

	\begin{figure}[tb]
		\centering
		\includegraphics[width=0.5\linewidth]{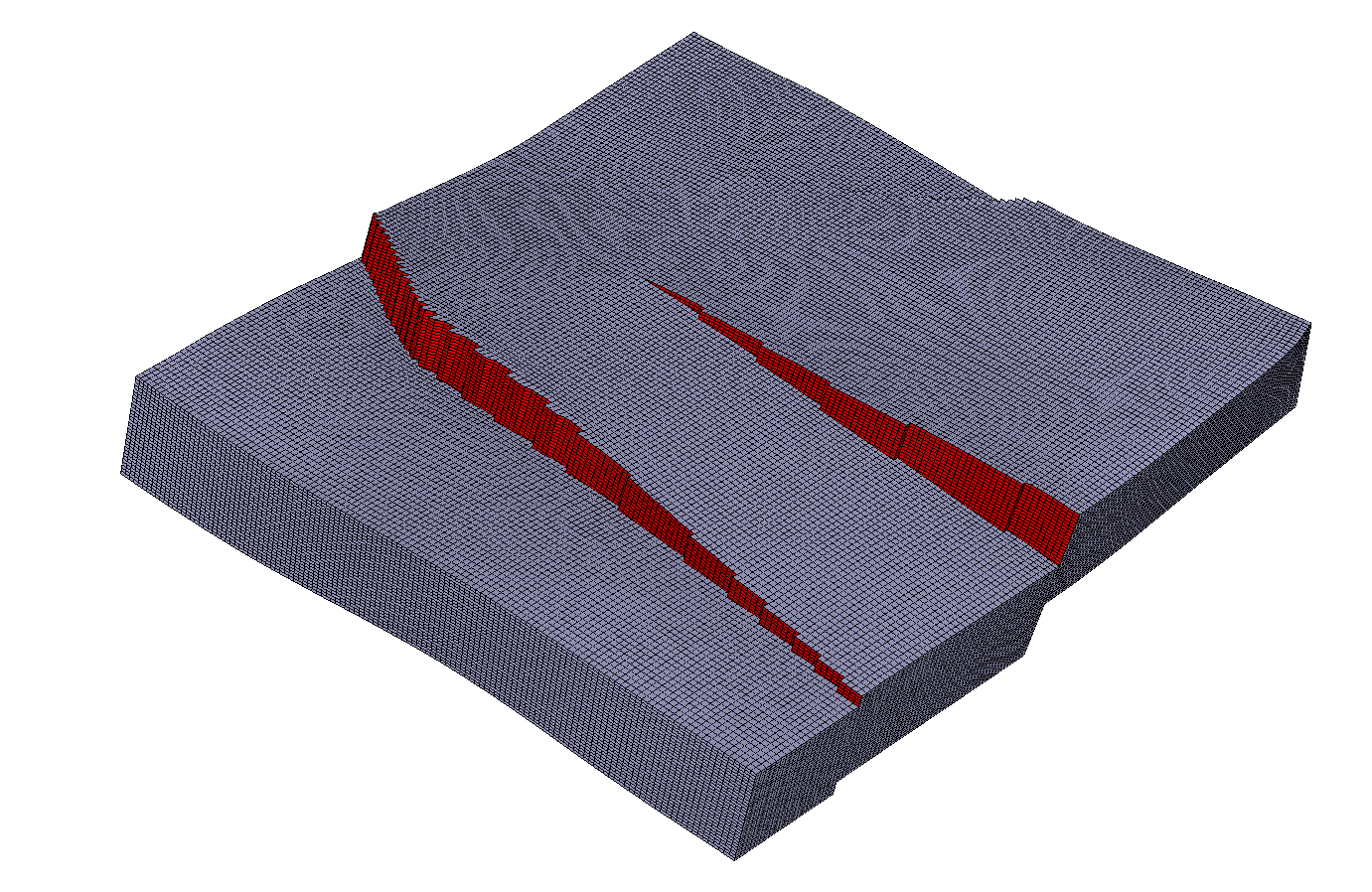}
		\caption{ \lund model mesh  (\numproduct{128 x 128 x 32} cells, three faults).\label{Fig_Lundi}}
	\end{figure}
Figure \ref{Fig_Lundi} provides an overview of the model hexahedral mesh underlying all \lund models. This mesh bears a geological morphology similar to SPE10 dataset 2 (quoting): a ``\emph{simple geometry, with no top structure or faults}''. It mainly differs in its lengths in each dimension (chosen as powers of two) and the addition of faults, which are challenging for upscaling/upgridding, multiscale decomposition, mesh compression (as vertices are not conform) and flow simulation (as faults affect fluid displacement).
	
\begin{figure}[tb]
		\centering
		\includegraphics[width=0.5\linewidth]{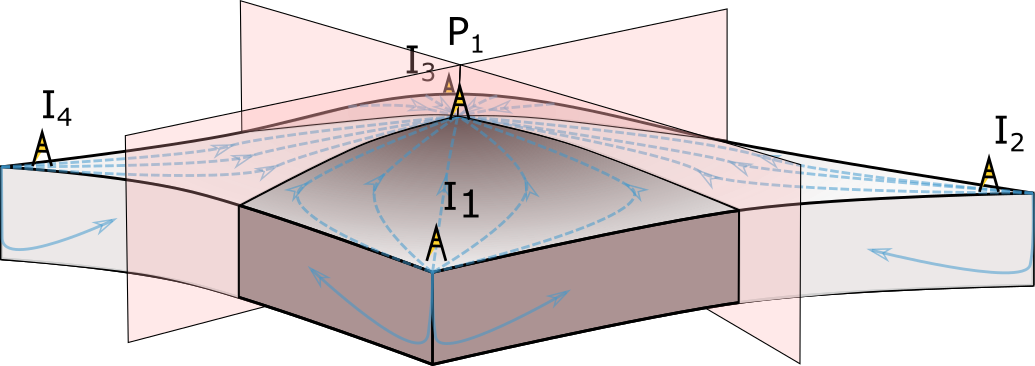} 
		\caption{Reservoir engineering: an injection/production system.  
		In this quarter five-spot configuration model \cite{Lie_K_2019_book_introduction_rsumgnuo},  waterflows (depicted by arrowed blue lines) are produced via injection from four wells ($\{I_i\}$) up to one central producer well $P_1$, located on the highest point. \lund represents one quarter of this reservoir model (highlighted in brown and limited by vertical reddish planes), including the wells $I_1$ and $P_1$. \label{fig_QuarterFiveSpotModel}}
			\end{figure}

	The topography of \lund models stems from a realistic   reservoir engineering case. It forms one quarter of an anticline structure (Figure \ref{fig_QuarterFiveSpotModel}), common in hydrocarbon trap reservoir study. The highest point ($P_1$) corresponds to the top of the anticline (\SI{3360}{m} depth). The opposite corner is situated \SI{50}{m} below, on the same horizon.
	
  \lund model mesh contains three continuous vertical stair-step faults. Two are apparent in red within Figure \ref{Fig_Lundi}, the third one bulging from the top-right side. They are not aligned along grid axes and  possess different offsets to emulate mildly complex environments. Its structure is composed of
	\numproduct{128 x 128 x 32} 
	cells to allow reasonable simulation times.  The average cell size represents a volume of size
	\qtyproduct[round-mode=places,round-precision=2]{1.7 x 1.7 x 0.95}{\metre},
	 which is common in sedimentary geology for modeling horizontal fine deposits of geologic material over the years. The  numbers of cells	(\numlist{128;128; 32})  in each dimension are powers of two (${2^7;2^7; 2^5}$). This choice allows to implement the most standard dyadic computations, subsampling or decompositions, to better benchmark compression methods. This choice allows to scale the mesh dimensions by   five scales, down to LEGO\textsuperscript{\textregistered{}} brick sizes. Note that size reduction by one or two dyadic scales often suffices. Therefore in practice,  non-dyadic dimensions may be handled by only padding cells to the next even or quadruple integer, or using activity labels. 
	
\subsection{\lund model properties} \label{sec_model_description_properties}

\begin{figure*}[htbp]
	\centering
	\subfigure[scale]{\label{Fig_scale_poro}\includegraphics[height=0.12\textheight]{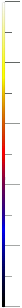}}
	\subfigure[\nsZ]{\label{Fig_nsZ_poro}\includegraphics[height=0.12\textheight]{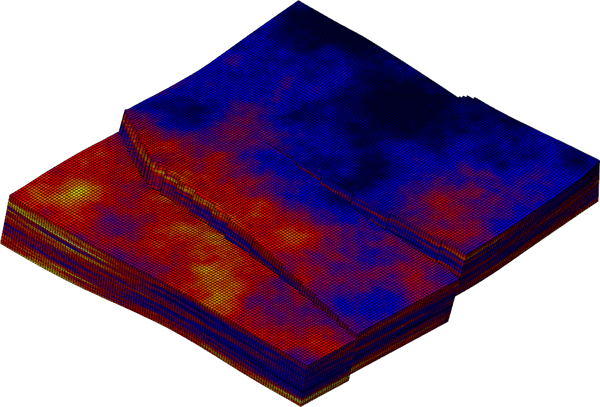}}
	\subfigure[\nsO]{\label{Fig_nsO_poro}\includegraphics[height=0.12\textheight]{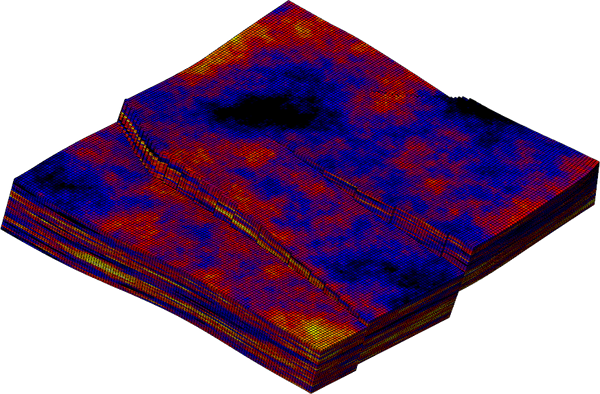}}
	\subfigure[\nsA]{\label{Fig_nsA_poro}\includegraphics[height=0.12\textheight]{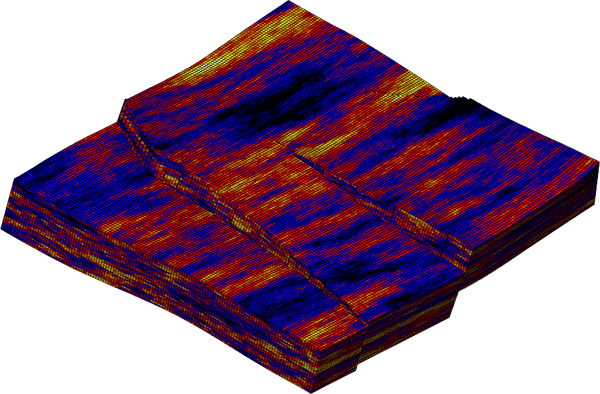}}
	\subfigure[\flv]{\label{Fig_flv_poro}\includegraphics[height=0.12\textheight]{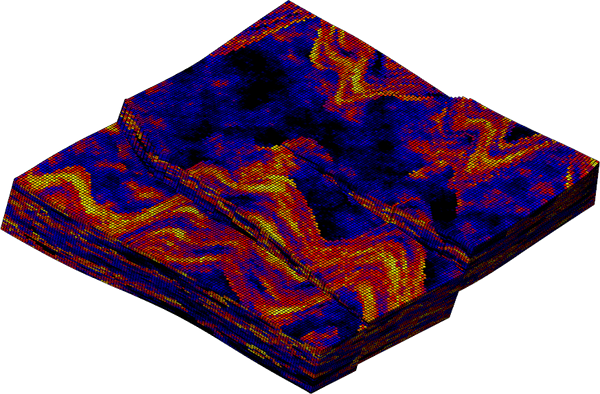}}\\
	\subfigure[scale]{\label{Fig_scale_perm}\includegraphics[height=0.12\textheight]{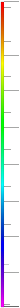}}
	\subfigure[\nsZ]{\label{Fig_nsZ_perm}\includegraphics[height=0.12\textheight]{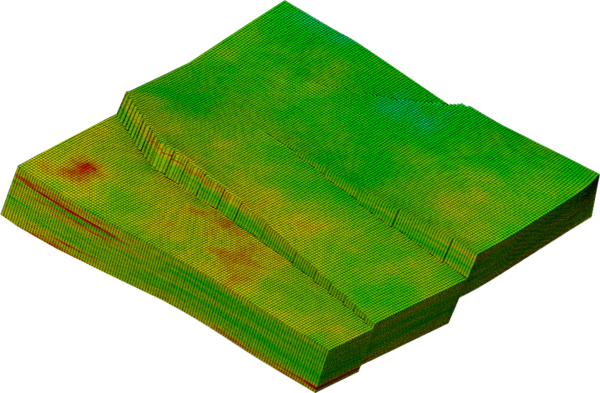}}
	\subfigure[\nsO]{\label{Fig_nsO_perm}\includegraphics[height=0.12\textheight]{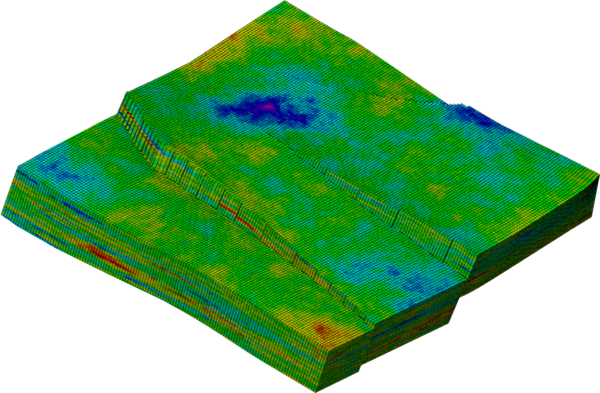}}
	\subfigure[\nsA]{\label{Fig_nsA_perm}\includegraphics[height=0.12\textheight]{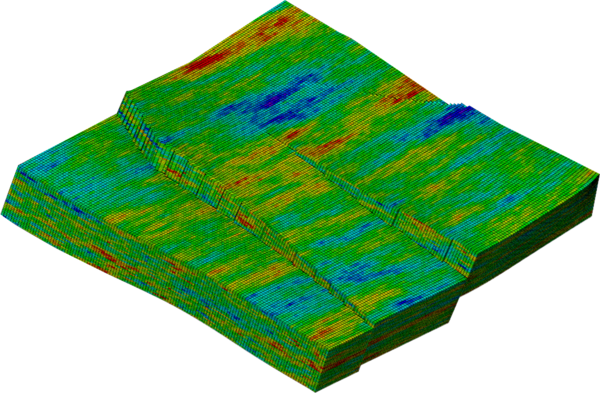}}
	\subfigure[\flv]{\label{Fig_flv_perm}\includegraphics[height=0.12\textheight]{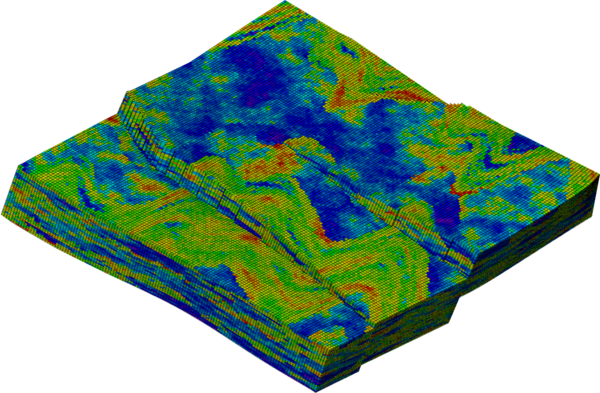}}
	\caption{Four geological environments supplied in \lund dataset: nearshore$_0$, nearshore$_1$, nearshore$_a$, and fluvial. Porosity (top) and permeability (bottom) properties range from \numrange[round-mode=places,round-precision=1]{0}{0.5} (as a unit fraction)  and \SIrange[scientific-notation = engineering]{0.0007}{20000}{\milli\darcy}, respectively.\label{Fig_LundiProperties}}	
\end{figure*}

The mesh is enhanced by two continuous petrophysical properties, porosity and permeability, required for the simulation benchmark, partly presented in \cite{Bouard_L_2021_p-coresa_etude_cicpvdsg}.
	The spatial distributions of those properties are inspired by two geological formations in \cite{Christie_M_2001_j-spe-reserv-eval-eng_tenth_specspcut}: \emph{Ness}\footnote{\url{https://data.bgs.ac.uk/id/Lexicon/NamedRockUnit/NESS}.} and \emph{Tarbert}\footnote{\url{https://data.bgs.ac.uk/id/Lexicon/NamedRockUnit/TARB}.} 
	Note that we do not consider here rock types: though they are important in overall compression schemes \cite{Peyrot_J_2019_j-computat-geosci_hexashrink_pdgpcshmg}, they were not required for our flow simulation purpose. As there is no obvious mapping from one geological object to another, we draw four different stochastic realizations to emulate four distinct environments, from homogeneous to  anisotropic, which are displayed in color scales in Figure \ref{Fig_LundiProperties}. 

The three first  correspond to prograding nearshore environments (\emph{Tarbert} formation) with  smooth property variations: \emph{\nsZ} and  \emph{\nsO}  have been generated by an isotropic distribution with different ranges of dependence, while \emph{\nsA} exhibits more  anisotropy. The fourth ``fluvial'' model   (\emph{Upper Ness} formation) exhibits sharper contrasts, with distinctive heterogeneous geological objects. This discrepancy between environments emulates a wide range of petrophysical system behaviors.  \lund meshes are available at  \href{10.5281/zenodo.14641959}{10.5281/zenodo.14641959}.

The conception of the initial common grid, the inclusion of faults and property filling have been performed with Paradigm\texttrademark{} 3D geological modeling software GOCAD (formerly known as GeOCAD, Geological Objects Computer-Aided Design; now SKUA)\footnote{\url{https://www.aspentech.com/en/products/sse/aspen-skua}} and the MATLAB Reservoir Simulation Toolbox (MRST)  \cite{Lie_K_2019_book_introduction_rsumgnuo}.

\section{Simulation settings}\label{sec_simulation_settings}
The following provides ancillary data, adapted from SPE10, so that \lund can be simulated through a Minimal Reproducible Example  (MRE) workflow. Precise outcomes may of course depend on alternative software choices and expert decisions.
\subsection{Global reservoir characteristics}\label{sec_global_reservoir_characteristics}
As for the reservoir model, the rock compressibility  is set to
	\SI{1e-6}{\per\bar}
	 and the reservoir pressure is set to 
	 	\SI{200}{\bar} 
	 	at the water-oil contact, fixed at a depth of
	 		\SI{3410}{\m}.
	 	 Finally, the reservoir temperature is set to \SI{60}{\celsius}.
	 	 
The simulation workflow is backed on the test case of a so-called black-oil model \cite{Abraham_F_2019_j-vis-comput_multiresolution_vmborm}.	
It consists of  two liquid phases:  water and dead oil  (with no gas dissolved). We introduce the 
Formation Volume Factor (FVF) quantity: ratio of  volumes occupied by a fluid at  reservoir conditions  versus  surface conditions. Quantities below are again borrowed from SPE10, and recalled for completeness. 
For water, viscosity pressure, density and FVF are computed by correlation from the reservoir simulator Pumaflow\textsuperscript{\textregistered{}}\footnote{\url{https://www.beicip.com/pumaflow}}.
For oil, some quantities are given by tabulations. The viscosity pressure (in centipoise, cP)  is computed from Table \ref{Tab_OilViscovityPressure}, the density is set to 	\SI{1}{\kilo\gram\per\meter^3},
and the oil FVF (B\textsubscript{o}) is tabulated in  Table \ref{Tab_OilVolumeFactorPressure}.

\begin{table*}[htbp]%
\centering %
\caption{Oil viscosity in  centipoise (cP), tabulated as a  function of pressure.\label{Tab_OilViscovityPressure}}
\begin{tabular}{ll}
\toprule
	\textbf{Pressure (bar)}  & \textbf{Viscosity (cP)} \\
\midrule
	50 &  2.85 \\
	200 &  2.99 \\
\bottomrule
\end{tabular}
\end{table*}


\begin{table*}[htbp]%
\centering %
\caption{Oil Formation Volume Factor (FVF) B\textsubscript{o}, tabulated as a function of pressure.\label{Tab_OilVolumeFactorPressure}}
\begin{tabular}{ll}
\toprule
	\textbf{Pressure (bar)}  & \textbf{B\textsubscript{o}} \\
\midrule
	50 &  1.05 \\
			200 &  1.02 \\
			500 & 1.01 \\
\bottomrule
\end{tabular}
\end{table*}


We now turn to water/oil mixture characteristics with relative permeabilities for water ($\mathrm{S\textsubscript{w}}$) and oil ($\mathrm{S\textsubscript{o}}$), respectively.  Given that the latter is obtained  from the former by $\mathrm{S\textsubscript{o}}  = 1- \mathrm{S\textsubscript{w}}$, we tabulate relatives permeability curves in Table 	\ref{Tab_RelativePermeability}, for water ($\mathrm{Kr\textsubscript{w}}$) and oil ($\mathrm{Kr\textsubscript{o}}$). Here, the irreducible water saturation is $\mathrm{S\textsubscript{wi}}=0.2$ (Table 	\ref{Tab_RelativePermeability}, top of first column) and the residual oil saturation is $\mathrm{S\textsubscript{or}}=0.2$ (complement to the $\mathrm{S\textsubscript{w}}$ given in Table \ref{Tab_RelativePermeability}, bottom of first column).

\begin{table*}[!t]%
\centering %
\caption{Relative permeability curves tabulation for water ($\mathrm{Kr\textsubscript{w}}$) and oil ($\mathrm{Kr\textsubscript{o}}$) as a function of water saturation ($\mathrm{S\textsubscript{w}}$).
\label{Tab_RelativePermeability}}
\begin{tabular}{lll}
\toprule
	\textbf{$\mathrm{S\textsubscript{w}}$} & \textbf{$\mathrm{Kr\textsubscript{w}}$} & \textbf{$\mathrm{Kr\textsubscript{o}}$} \\ 
\midrule
	0.200$^\dag$ &0.0000& 1.0000 \\
			0.250 &0.0069& 0.8403 \\
			0.300 &0.0278& 0.6944 \\
			0.350 &0.0625& 0.5625 \\
			0.400 &0.1111& 0.4444 \\
			0.450 &0.1736& 0.3403 \\
			0.500 &0.2500& 0.2500 \\
			0.550 &0.3403& 0.1736 \\
			0.600 &0.4444& 0.1111 \\
			0.650 &0.5625& 0.0625 \\
			0.700 &0.6944& 0.0278 \\
			0.750 &0.8403& 0.0069 \\
		0.800$^\ddag$ &1.0000& 0.0000 \\
\bottomrule
\end{tabular}
\begin{tablenotes}
\item[$\dag$] Irreducible water saturation value ($\mathrm{S\textsubscript{wi}}$).
\item[$\ddag$] Residual oil saturation value ($\mathrm{S\textsubscript{or}}$).
\end{tablenotes}
\end{table*}

\subsection{Application to fluid production}\label{sec_fluid_production}

	We finally present a typical two-phase flow simulated on \lund (full resolution, nearshore$_0$ environment).  Initially, two phases in the reservoir are horizontally stratified, with oil above water. The two wells are drilled in the whole depth of the reservoir. At $t=0$,  water is injected by $I_1$ in the lower part of the reservoir (Figure \ref{fig_QuarterFiveSpotModel}).  The water pressure pushes the oil through the reservoir up to the producer $P_1$ (distant from \SI{300}{m}). 
	Injector pressure and producer rate remain constant,  respectively set at \SI{300}{\bar} and \SI{300}{m^3} per day.  
	
	One valuable indicator in oil production to determine field exploitation is the estimated water cut, i.e., the ratio between water and total liquid volumes, at the producer well. It is regularly recorded over a period of time expressed in days (Figure {\ref{Fig_QuarterFiveSpot_Watercut}). The inflection point (red point on the curve) is the water breakthrough, which denotes the water arrival at $P_1$. From that instant, the extracted liquid contains more and more water. To avoid expensive post-processing and optimize the exploitation configuration, reservoir engineers aim to delay this instant. Simulation is a powerful tool to estimate such predicted water cut curves. To determine the best exploitation configuration, many simulations with varying parameters can thus be run until satisfaction, involving very long computation times. We expect to evaluate the positive impact of resolution and precision variations (on computational time) on the latter.

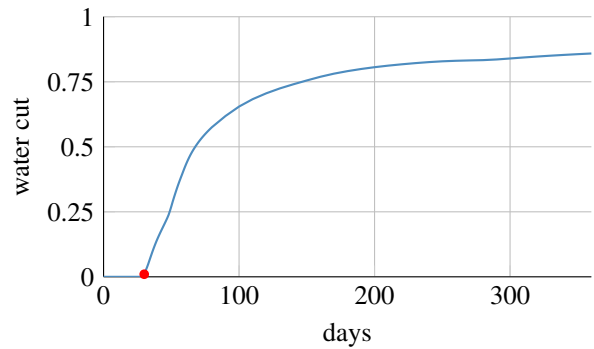
\begin{figure}[htbp]
	\centering
		\begin{tikzpicture}
		\definecolor{bluegray}{rgb}{0.30, 0.55, 0.75}
		\begin{axis}[xmode=normal,ymode=normal,
		width=8cm,  height=5cm,
		axis y line*=left,
		axis x line*=bottom,
		ytick={0,0.25,...,1},xtick={0,100,...,300},
		grid,
		xlabel ={days},
		ylabel={water cut},
		xmin=0, xmax=360,
		ymin=0, ymax=1, 
		]
		\addplot[color=bluegray,thick] coordinates{(0,0)(1,0)(2,0)(3,0)(4,0)(5,0)(6,0)(7,0)(8,0)(9,0)(10,0)(11,0)(12,0)(13,0)(14,0)(15,0)(16,0)(17,0)(18,0)(19,0)(20,0)(21,0)(22,0)(23,0)(24,0)(25,0)(26,0)(27,0)(28,0)(29,0)(30,9.821302e-03)(31,2.348330e-02)(32,3.566801e-02)(33,4.988481e-02)(34,6.538830e-02)(35,8.064744e-02)(36,9.573865e-02)(37,1.100073e-01)(38,1.236561e-01)(39,1.366514e-01)(40,1.488025e-01)(41,1.603689e-01)(42,1.715230e-01)(43,1.823665e-01)(44,1.931034e-01)(45,2.038341e-01)(46,2.147645e-01)(47,2.259806e-01)(48,2.381895e-01)(49,2.538089e-01)(50,2.705831e-01)(51,2.886781e-01)(52,3.056738e-01)(53,3.220320e-01)(54,3.378018e-01)(55,3.529809e-01)(56,3.673767e-01)(57,3.811389e-01)(58,3.946305e-01)(59,4.084584e-01)(60,4.218233e-01)(61,4.343728e-01)(62,4.462050e-01)(63,4.573320e-01)(64,4.676945e-01)(65,4.773061e-01)(66,4.862838e-01)(67,4.947211e-01)(68,5.026650e-01)(69,5.102103e-01)(70,5.174560e-01)(71,5.243418e-01)(72,5.308987e-01)(73,5.371683e-01)(74,5.431957e-01)(75,5.489944e-01)(76,5.545981e-01)(77,5.600404e-01)(78,5.653232e-01)(79,5.704344e-01)(80,5.753854e-01)(9.004167e+01,6.185195e-01)(1.000417e+02,6.547598e-01)(1.100417e+02,6.830940e-01)(1.200417e+02,7.055570e-01)(1.300417e+02,7.243308e-01)(1.400417e+02,7.402824e-01)(1.500417e+02,7.551094e-01)(1.600417e+02,7.692191e-01)(1.700417e+02,7.811622e-01)(1.800417e+02,7.908738e-01)(1.900417e+02,7.990628e-01)(2.000417e+02,8.060578e-01)(2.100417e+02,8.120007e-01)(2.200417e+02,8.170962e-01)(2.300417e+02,8.216335e-01)(2.400417e+02,8.255616e-01)(2.500417e+02,8.286396e-01)(2.600417e+02,8.306564e-01)(2.700417e+02,8.320229e-01)(2.800417e+02,8.331606e-01)(2.900417e+02,8.357197e-01)(3.000417e+02,8.394942e-01)(310,8.433689e-01)(320,8.469872e-01)(330,8.502931e-01)(340,8.533257e-01)(350,8.561244e-01)(360,8.587030e-01)};

		\addplot[color=red, mark=*, thick, mark options={scale=0.7}] coordinates{(30,0.01)};
		
		%
		\end{axis}
		\end{tikzpicture}
	\caption{Two-phase flow water cut simulated on \lund (full resolution, nearshore$_0$ environment) and measured  at $P_1$ according to the configuration of Figure \ref{fig_QuarterFiveSpotModel} (quarter five-spot model). Two-phase flow  simulated water cut is measured  at $P_1$. The red dot indicates the time (in days) of the water breakthrough, i.e. when water starts being produced at well $P_1$.\label{Fig_QuarterFiveSpot_Watercut}} 
\end{figure}

		\section{Data format and access}\label{sec_data_format} 
The four \lund models presented in this article (one per environment) are provided as ``Grid Eclipse" (GRDECL) data, a \emph{de facto} standard for grids with hexahedral cells,  developed by Schumberger for the ECLIPSE\texttrademark{} Reservoir Simulator\footnote{\url{https://www.software.slb.com/products/eclipse}}. They are available  at Zenodo\footnote{\url{https://doi.org/10.5281/zenodo.14641958}} and from author website\footnote{\url{http://www.laurent-duval.eu/opus-lundisim.html}} \cite{Duval_L_2025_zenodo_lundisim_odmmmpppispe10cgdsa}.  Lower resolutions of \lund, produced by  \hs, are also available.
		
	 \lund  illustrations from Figures \ref{Fig_LundiProperties} and \ref{Fig_Lundi} were made with \emph{ResInsight}\footnote{\url{https://resinsight.org}} (v.2023.06\footnote{\url{https://github.com/OPM/ResInsight}}), an open source cross-platform 3D visualization and post-processing tool for reservoir models and simulations (developed in Python, available for Windows and Linux). Other Python libraries support GRDECL format, for instance \emph{PyGRDECL}\footnote{\url{https://github.com/BinWang0213/PyGRDECL}} or \emph{XTGeo}\footnote{\url{https://pypi.org/project/xtgeo}}, and can be also used for visualization or other processings.
	
	\section{Potential dataset use/reuse}\label{sec_data_reuse}

Inspired by the SPE10 reservoir simulation challenge \cite{Christie_M_2001_j-spe-reserv-eval-eng_tenth_specspcut}, \lund with its different environments are primarily meant for evaluating the performance of lossy or lossless compression algorithms with respect to reservoir modeling and simulation. Openly-shared models are scarce in reservoir geoscience and engineering. \lund  serves other purposes as well.

It can be used to test more geologically-oriented upscaling methods and their  reliability regarding information loss, through quality indicators \cite{Preux_C_2014_j-ogst_use_qirilwpu}. While initially developed for hydrocarbons, our approach may  be used for  sustainable challenges, for instance geothermy,  hydrogen (H\textsubscript{2})  or carbon dioxide (CO\textsubscript{2} \cite{Alumbaugh_D_2023_j-geosci-data-j_kimberlina_smdco2mi})  storage projects. Note that the Society of Petroleum Engineers has  released a call on the 11th SPE challenge for safe and efficient implementation of geological carbon storage.

Being complex volume meshes, \lund models can be used to benchmark  scientific data compression algorithms (either on the 3D volume data or meshes). They are also adapted to investigate
the impact of reduced data precision  \cite{Moreland_K_2022_tr_exploitation_drv} or resolution change on pure objective metrics (for instance in a context of mesh visualization, storage or checkpoint restart), but also on faithfulness of any simulation. 

As for precision, current practice favors the IEEE 754 floating-point format --- in double, quadruple or even octuple precision \cite{Gladman_B_2024_tr_accuracy_mfsddeqp} --- to ensure both accuracy and simplicity of data management.  As a result, some data fields are represented, stored and transferred with an excessive number of bits \cite{Walters_M_2023_j-geosci-model-dev_impact_aedpceascmaqm}. Plus, it is being recognized that for a given simulation workflow,  quantities from an homogeneous data field  may possess widely different statistical distributions, in which distinct scales of magnitude are associated to different spread/precision/impact. For instance, a permeability value of zero or below \SI{50}{\milli\darcy}  means ``no to meaningless'' water flows (rocks working  as ``seals''), while  values greater by orders of magnitude (over \SI{10000}{\milli\darcy}) may yield ``full permeation''. As a consequence, a fine precision for small permeabilities is meaningful, when higher permeability values would not affect results when changed by $ \SI{\pm20}{\percent}$. As HPC sparks interest on so-called next-generation arithmetic (such as UNUM or POSIT formats \cite{DeDinechin_F_2019_p-conga_posits_gbu,Lindstrom_P_2022_p-conga_multiposits_ucrn,Kneusel_T_2024_inbook_posits}),  with so few simulation tools  already adapted to such hybrid data formats, it is important to be able to emulate them on shared and representative dataset with  minimal, reproducible  examples  of workflows.}

As for  resolution, with edge computing, or the necessity sometimes to assess  crude estimations in real time \cite{Sicat_R_2023_j-ieee-trans-visual-comput-graph_real-time_vlsgmnfpld} on low-power devices using  cloud resources, it becomes increasing important to provide users data with adapted granularity. One straightforward scheme  consists in sharing the original data source as well as several lower-resolution versions, either with pyramid schemes \cite{Ceballos_L_2021_incoll_interactive_dp3dvse} or with embedded multiresolution mechanisms, for instance with  wavelets \cite{Christophe_E_2008_j-ieee-tip_hyperspectral_icaspihtezwa3dwc,Jacques_L_2011_j-sp_panorama_mgrisdfs}  as in  \cite{Peyrot_J_2019_j-computat-geosci_hexashrink_pdgpcshmg}. For this reason, we provide  \lund models with their associated lower resolution representations. The latter may also be probed with varying precision, as mentioned previously. 
Evaluating the combined  impact of resolution and precision is briefly evoked in \cite{Bouard_L_2021_p-coresa_etude_cicpvdsg}, and  the  topic for a forthcoming  companion paper. 
Additional reuse cases reside in combining simulation and compression with machine learning or artificial intelligence tools,  which are being used more intensively in simulation \cite{Glaws_A_2020_j-phys-rev-fluids_deep_lisdcltfs}.

Future research  may be interested in larger-size models than those we share here. By providing here the main  ingredients and philosophy used to build \lund models, we hope they will help in creating novel meshes along our open methodological guidelines.

		\section{Conclusion}\label{sec_data_conclusion}
		
		A couple of years ago, due to the lack of openly shared heterogeneous and realistic geoscience data  to study influence of compression on simulation workflows,  we designed our own models,  inspired by the SPE10 challenge. For other researchers on this field to overcome this pitfall, we  now share our models named \lund to the scientific community in the FAIR spirit. Based on a typical geoscientific mesh containing several faults, and two formations proposed in SPE10, we generated four models with distinct environments, including porosity and permeability information. Thanks to the multiresolution \hs framework, our dataset also includes lower-resolution versions of each model (mesh and attributes), with consistent fault preservation whatever the level of decomposition. We hope that this dataset will be useful to other geoscience researchers in taking their projects forward.
		

\bmsection*{Acknowledgments}
The research presented was mainly performed during the PhD thesis of Lauriane Bouard, following the post-doctoral position  of Jean-Luc Peyrot and the internship of Lena\"ic Chizat. The authors are grateful to IFP Energies nouvelles for the permission to share  \lund models. They acknowledge the support of AIR (Action IUT Recherche) of IUT Côte d'Azur. They thank Christophe Latry, Carole Thiébaut (CNES), Laurent Astart, Nadine Couëdel, Frédéric Douarche, Thomas Guignon (IFP Energies nouvelles), Corinne Maihles (TéSA) and Marc Antonini (UniCA, CNRS) for their support.

\bmsection*{Financial disclosure}
None reported.

\bmsection*{Conflict of interest}
None.

\bmsection*{Supporting information}

A  companion paper \cite{Duval_L_2025_PREPRINT_how_drrpcpdafsblm} about the impact of reduced resolution/precision and companding with HexaShrink on the proposed simulation is planned (partly presented in \cite{Bouard_L_2021_p-coresa_etude_cicpvdsg}).

\end{document}